\documentclass[showpacs,twocolumn,floatfix,superscriptaddress]{revtex4}
\input{epsf}

\newcommand{\bq}{\begin{equation}}
\newcommand{\ee}{\end{equation}}
\newcommand{\fr}[2]{\frac{#1}{#2}}
\newcommand{\eps}{\varepsilon}

\begin{document}
\draft

\title{ Spin Effects and Transport in Quantum Dots
with overlapping Resonances}
\author{P.G.Silvestrov}
 \affiliation{Budker Institute of Nuclear Physics, 630090
Novosibirsk, Russia}
 \affiliation{Weizmann Institute of science, Rehovot 76100,
Israel}
 \affiliation{Instituut-Lorentz, Universiteit Leiden, P.O. Box 9506, 2300
RA Leiden, The Netherlands}

\author{Y.Imry}
 \affiliation{Weizmann Institute of science, Rehovot 76100,
Israel}

\date{today}
\begin{abstract}

The role of spin is investigated in the transport through a
quantum dot with two overlapping resonances (one having
a width larger than the level separation and the other
very narrow, cf. Silvestrov and Imry, Phys. Rev. Lett. {\bf 85},
2565 (2000)). For a series
of consecutive charging resonances, one electron from the leads
populates one and the same broad level in the dot. Moreover,
there is the tendency to occupy the same level also by the
second electron within the same resonance. This second electron
is taken from the narrow levels in the dot. The narrow levels
are populated (and broad level is depopulated) via  sharp
rearrangements of the electronic configuration in the Coulomb
blockade valleys. Possible experimental manifestations of
this scenario are considered. Among these there are sharp features in
the valleys and in the Mixed Valence regime and an unusual
Kondo effect.

\end{abstract}


\maketitle


\section{ Introduction.}

In this paper we consider the transmission through a
multilevel quantum dot~\cite{LKCM} having only one
broad level well coupled
to the leads. Such a model has been suggested in Ref.~\cite{SI}
in order to provide an explanation for the behavior of
the transmission phase through a quantum dot observed in the experiment
Ref.~\cite{Heiblum2}.
Within that model, electrons are transferred from the leads
to the broad level in the quantum dot within the charging peaks,
and are then transferred to and "stored"  in the narrow levels
via sharp transitions in-between
peaks. Thus the charging
(or conductance) peaks are very similar to each other,
in the behaviors of both the conductance and the transmission phase.
In the present paper we treat the effects associated with the
spin-degeneracy of the levels for such QD-s.

Many of the sharp
rearrangements of the electronic configuration in the QD, which
we will consider, take place due to spin, or are sufficiently
modified compared to the spinless case to make the discussion
of spin rather interesting. The strong coupling of one level to the
leads leads now to the tendency to have this ``valence'' level
either doubly occupied or completely empty. However, the total number
of electrons in the dot may change only by one at any charging
resonance. In particular the resolution of this formal contradiction
leads to prediction of singular behavior of the conductance just at
the top of charging resonance.

The peculiar temperature (bias) dependence of the Kondo
effect~\cite{KondoT1,KondoT2,KondoT3}, which takes
place in the transport through the quantum dot with nonzero spin, may
facilitate the experimental verification of our predictions.
Moreover, in the experiments (see e.g.~\cite{Kondo1,Kondo2,Kondo3})
designed to observe the Kondo effect, in order to
measure the conductance in the Coulomb
blockade valley the dot is strongly coupled to the leads.
Thus it may naturally happen that the widths of some accidentally
close levels will exceed the inter level energy difference.
However, the new effects which we consider in this paper either
take place in the mixed valence regime, or should be seen already
at the temperature sufficiently higher than the Kondo temperature.

In general, we consider the situation when the transport through
the quantum dot has well defined resonance structure with pronounced peaks
and Coulomb blockade valleys. However, the coupling to the leads of the single
broad level is already strong enough to change the usual
formation of the many-electron ground state of the dot via
the consecutive occupation of single particle levels. The
energetics of such a system is described in the following section.
In Section~III we consider the conductance of
such a quantum dot in the limit of vanishing width of all narrow ``spectator''
levels. In the Sec.~IV we investigate the smearing of the sharp
features of the conductance predicted in the Sec.~III due to
the finite temperature or width of narrow level.
Following the experimental tendency for miniaturisation of the QD-s
we mainly consider only a two-level dot (with one narrow level and
one having the width exceeding the interlevel energy spacing).
However, in Section V we consider the effect of spin for the
transport in the quantum dot having many narrow and one broad level with
$\Gamma\gg\Delta$ (with $\Gamma$ being the width of the broad level
and $\Delta$ the interlevel spacing). Discussion and conclusions
are given in the Section VI.

\section{ The ground state energy.}

To model the quantum dot we use the tunnelling Hamiltonian in the constant
interaction~($U$) approximation
\begin{eqnarray}\label{Ham}
&\,&H=\sum_i \eps_i a^+_i a_i + U\sum_{i<j} a^+_i a_i  a^+_j a_j
+ \\
&\,&   \sum_k\eps^L(k) b^{L+}_k b^L_k +
\sum_{k,j}[t^L_j a^+_j b^L_k +H.c.]
+ L\leftrightarrow R
\ . \nonumber
\end{eqnarray}
Here $a$ and $b^{L(R)}$ are the annihilation operators for the
electrons in the dot and in the left(right)lead, respectively. We will use
$\eps(k)=k^2/2m-E_F$ and will not introduce any $k$ dependence
of the tunnelling matrix elements $t_j$.
All summations in
eq.~(\ref{Ham}) include also the summation over spin. Only one
level, $\eps_1$, is well coupled to the leads, having the width
\bq\label{Gamma}
\Gamma\equiv \Gamma_1 = 2\pi\sum_{i=L,R} |t_1^i|^2
{dn^i}/{d\eps} > \Delta \ .
\ee
We take $\Gamma\ll U$.
Since we have in mind the experiments with really very small quantum dots,
let us consider only two
levels in the dot (e.g.  having accidentally close energies).
The generalisation to many narrow levels will be considered in the
Sec.~V. We assume that coupling between the dot and the gate electrode
is pure capacitive.
Then the levels
flow uniformly with the gate voltage  $d\eps_i/dV=const$ and
without a loss of generality we may put
\bq\label{gate}
\eps_1=-V \ \ , \ \ \eps_2 =-V +\Delta
\ .
\ee
There are four charging resonances of the conductance $G(V)$ at
$V\approx 0,U,2U$ and $3U$ (see for the review on the Coulomb blockade
in quantum dots e.g. the ref.~\cite{Coulomb}).

Our first aim will be to find the ground state of the
system~(\ref{Ham}) at different values of $V$. In the limit
$\Gamma_2\rightarrow 0$ the number of electrons on the level $2$
is a good quantum number. Let us denote by $E^{(0)}, E^{(1)},
E^{(2)}$ the total energy of the lowest state of the
quantum dot interacting
with the leads, with the narrow level populated by, respectively,
$0,1$ and $2$ electrons
(more precisely, $E^{(i)}$ is defined as the eigenenergy of the
Hamiltonian~(\ref{Ham}) minus the trivial energy of the
electrons in the leads $\sum\eps(k)$). The functions
$E^{(i)}(V)$ evolve smoothly with the gate voltage and the
(averaged) occupation number of the broad level $1$ also changes
continuously. For example, the branch $E^{(0)}$ corresponds to an
empty level $1$ at $V<0$, singly occupied at $0<V<U$
and doubly occupied at $U<V$. For $t_2^{L,R}=0$ the
functions $E^{(i)}$ may cross at some values of $V$, which in
particular may lead to a sharp change of ground state.

With the use of perturbation theory in $t^{L,R}_1$ it is
easy to find $E^{(i)}$ far from the charging peaks.  Below the
first resonance (at $V\ll-\Gamma$) the true ground state is
evidently $E^{(0)}$. However, already here the virtual jumps
of the electrons from the wire to the level $1$ give rise to the correction
\begin{eqnarray}\label{3}
&&E^{(0)}=2\sum_{k<k_F}
\fr{|t^{L,R}|^2}{\eps(k)-\eps_1}
= \fr{- \Gamma}{\pi} \ln\left( \fr{4E_F}{\eps_1} \right) ,\\
&& E^{(1)}\approx \eps_2 \ , \ E^{(2)}\approx 2\eps_2 +U .\nonumber
\end{eqnarray}
The overall factor $2$ in $E^{(0)}$ accounts for the spin. It is clear
that $E^{(0)}$ in this region lies significantly below  $E^{(1)}$
and $E^{(2)}$ ($E^{(0)}$ is the true ground state).
The factor of four in the argument of the logarithm, is specific to
the dispersion of the conduction electrons we took. It does not appear in
the physical results.

When the (increasing) voltage crosses the region
$|V|\sim\Gamma$, the dot is charged by the
first electron. However, this electron may stay in the dot on
the level $1$ (described by $E^{(0)}$) or on the level $2$
($E^{(1)}$). Depending on what level is occupied the perturbation
theory gives, in this range of V
\begin{eqnarray}\label{33}
E^{(0)}&=& \eps_1 -\fr{\Gamma}{2\pi} \left\{ \ln\left(
\fr{4E_F}{|\eps_1|} \right) +\ln\left(
\fr{4E_F}{\eps_1+U} \right) \right\} \, ,\\
E^{(1)}&=& \eps_2 -\fr{\Gamma}{\pi} \ln\left(
\fr{4E_F}{\eps_1+U} \right) \ , \ E^{(2)}\approx 2\eps_2 +U
.\nonumber
\end{eqnarray}
The first logarithm in $E^{(0)}$ accounts for the virtual jumps of an
electron from the level $1$ in the dot to the wire. The other logarithms
correspond to  virtually adding the second electron to the dot
(having $\eps_1+U$ instead of $\eps_1$ in the denominator).
The two levels $E^{(0)}$ and $E^{(1)}$ cross at a gate voltage given
by \cite{SI}
\bq\label{8}
V=V^{I}=\fr{U}{\exp\{ -2\pi\Delta/\Gamma\} +1} \ .
\ee
This result is valid for both signs of $\Delta$. For $\Gamma
\gg|\Delta|$, eq.~(\ref{8}) reduces to $\eps_1\approx-U/2$.
At $V=V^{I}$ the electron in the dot jumps from the broad level to the
narrow one \cite{SI}. We will describe the consequences of such a ``jump''
for the transmission, below.

In the second valley, $U<V<2U$ the dot is charged already by two
electrons, which may populate  the two doubly degenerate levels in
the dot in different ways. Thus,
\begin{eqnarray}\label{33333}
E^{(0)}&=& 2\eps_1+U -\fr{\Gamma}{\pi}  \ln\left(
\fr{4E_F}{|\eps_1+U|} \right) \ , \\
E^{(1)}&=& \eps_1+\eps_2+U \nonumber\\
&& -\fr{\Gamma}{2\pi}
\left\{\ln\left(
\fr{4E_F}{|\eps_1+U|}\right) +\ln\left( \fr{4E_F}{\eps_1+2U}
\right)
\right\}\ , \nonumber\\
E^{(2)}&=& 2\eps_2+U -\fr{\Gamma}{\pi}  \ln\left(
\fr{4E_F}{\eps_1+2U} \right) \ . \nonumber
\end{eqnarray}
First of all, we see that just after the charging peak, at $V>U$
the true ground state is $E^{(0)}$. This is in contrast with the
situation at $V<U$, where the ground state was
$E^{(1)}$~(\ref{33}). Thus we may conclude, that within the
resonance not only is  one electron gradually transmitted  from the wire
to the level $1$ in the dot, but also a second electron is taken from
the narrow level $2$ to the broad one $1$ (see fig.~1). In the limit
$\Gamma_2\rightarrow 0$ this second ``transfer'' is abrupt and
takes place at some $V=W^I\approx U$ (we remind the reader that
we consider only the ground state energy in this section
and therefore $T\equiv 0$). This prediction of the
possibility to have sharp features within the charging resonance is
probably the main new result of this paper.

All the three energies (\ref{33333}) cross at the same value of
gate voltage (c.f. (\ref{8}))
\bq\label{88}
V^{II}=U+\fr{U}{\exp\{ -2\pi\Delta/\Gamma\} +1}
\ee
and the true ground state become $E^{(2)}$. Now already two
electrons jump together from the broad level to the narrow one.

At $V>V^{II}$ (\ref{88}) the occupation of the quantum dot proceeds in
a fashion which is almost symmetric under  particle--hole
replacement.  At the third peak the third electron is added to
the dot. Were the branch $E^{(2)}$  the stable one, this would have
been the uncoupled electron at the level $1$. However, $E^{(2)}$ and
$E^{(1)}$ cross at the top of the third peak (at
$V=W^{II}=W^I+U$) and the ground state for the first half of last
valley has a single unpaired electron at the narrow level. Finally, at
\bq\label{888}
V^{III}=2U+\fr{U}{\exp\{ -2\pi\Delta/\Gamma\} +1}
\ee
$E^{(2)}$ and $E^{(1)}$ cross and the broad level became singly
occupied again.  The fourth peak ($V\approx 3U$) completes the
charging of two levels in the quantum dot by four electrons.

\begin{figure}[t]
\epsfxsize=8.8cm
\epsffile{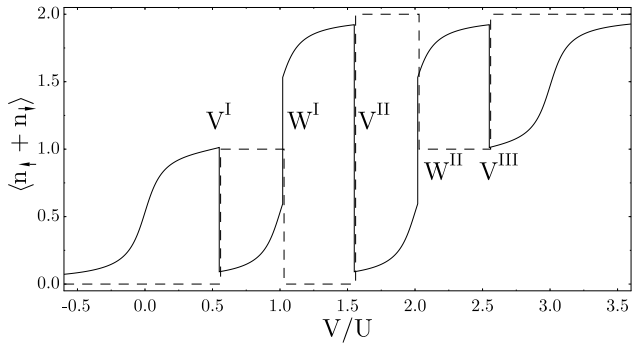}
\vglue 0.2cm
\medskip
\caption{Schematic drawing of the averaged occupation number
$\langle n_{\uparrow} + n_{\downarrow} \rangle$ for the broad
level $1$ (solid) and narrow level $2$ (dashed). The gate
voltage $V$ is measured in units of the charging energy $U$.
The charging resonances are at $V/U\approx 0,1,2,3$.}
\end{figure}

To summarise the discussion of this Section we show schematically
in fig.~1 the averaged occupation numbers of our two levels $\langle
n_{1\uparrow}+ n_{1\downarrow}\rangle$ and $\langle
n_{2\uparrow}+ n_{2\downarrow}\rangle$. The four charging resonances
correspond to $V/U\approx 0,1,2,3$. One may see on the figure many
abrupt changes in the population of both broad and narrow levels.
Unfortunately, the averaged occupation number of the given level
is not measured directly in the typical experiments with the quantum dots.
In order to make the connection with possible experiments we consider
in the following sections the transport properties of the quantum dot
described by the Hamiltonian eq.~(\ref{Ham}).

\section{ Conductance, leading order.}

The zero bias conductance $G$ of our
two-level quantum dot is shown schematically in  fig.~2. In the limit
of "invisible" level $2$ ($\Gamma_2 \rightarrow 0$),
we may introduce three conductances
$G^{(0,1,2)}$ corresponding to empty, singly- and
doubly-occupied narrow level (corresponding  to the three "ground state"
energies $E^{(0,1,2)}$ of the previous section). The role of
electrons at the level $2$
reduces in this case to simply raising  the current-transmitting
level $1$ via the Coulomb repulsion. Thus
\bq\label{rising}
G^{(0)}(V)=G^{(1)}(V+U)=G^{(2)}(V+2U) \ .
\ee
We have shown schematically the function $G^{(0)}$ on the same
fig.~2 (slightly offset vertically).
In the limit $\Gamma_2\ll\Gamma_1$ the curve for
two-level dot is obtained simply by cutting and horizontally
shifting the parts of the curve for single-level dot, in agreement
with eq. (\ref{rising}), as shown in the figure.
Thus the relations eq.~(\ref{rising}) allow one to describe the
singular behavior of the conductance even without the explicit
calculation of $G^{(0)}$. This result is particularly useful at
low temperatures close to the charging resonances and in the Kondo valleys,
where simple analytical formulas are not available.

\begin{figure}[t]
\epsfxsize=8.8cm
\epsffile{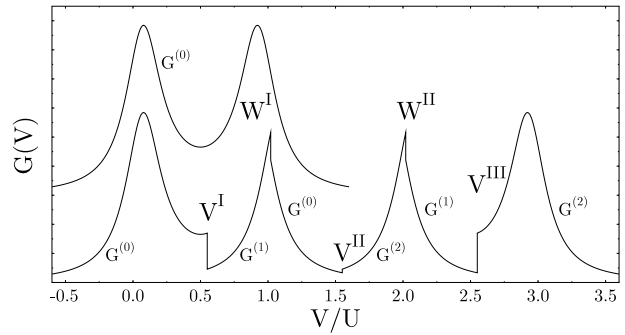}
\vglue 0.2cm
\medskip
\caption{The gate voltage dependence (schematic) of the conductance $G$.
The upper
curve shows $G^{(0)}$ in the case of only one (broad) level
in the quantum dot (vertically offset for clarity).
The lower curve depicts the conductance given by different
$G^{(n)}$ in various regimes.
Sharp features at two peaks and three valleys are seen.}
\end{figure}

Simple analytical expressions for the conductance may be found by means of
perturbation theory only far from the resonances.
In particular, below the first resonance
($V<0$) and above the second ($V>U$) one has for $G^{(0)}$
\bq\label{G1}
G^{(0)}=2G_Q
\fr{\Gamma_L\Gamma_R}{\eps^2} \ \ {\rm and} \
\ G^{(0)}=2G_Q
\fr{\Gamma_L\Gamma_R}{(\eps+U)^2} \ .
\ee
Here $G_Q=e^2/h$.
Because of
eq.~(\ref{G1}) being valid only far from the resonance, there is no
need to distinguish between $\eps_1$ and $\eps_2$. The first
formula here accounts for the virtual jump of the electron (with
any spin orientation) from the left lead first to the dot and
then to the right lead. For  $V>U$, first one of the $2$ electrons
jumps from the dot to the right lead, then another electron from the
left lead fills in its place in the quantum dot.

In the valley $0< V <U$
one electron stays at the level $1$ in the dot. Let it have e.g.
a spin up $\uparrow$.
(The total current is evidently the ``average'' of two equal currents
for the dot up~$\uparrow$ and the dot down~$\downarrow$).
We calculate the current via the transmission of electrons from the
left to the right lead.
There  are two contributions to the spin-conserving
current: either an electron with spin down $\downarrow$ goes from
the left lead to the dot and then to the right lead (energy denominator
$\eps+U$), or the  $\uparrow$ electron
from the quantum dot goes to the right lead  (energy denominator
$\eps$) and then  a $\uparrow$ electron
jumps to the dot from the left lead.  The {\em probabilities} of these
two processes should be added. Then, there are two spin-flip processes,
whose total current should be added to the spin-conserving one above:
either a $\downarrow$ electron goes from
the left lead to the dot (energy denominator $-(\eps+U)$
and then the $\uparrow$ electron goes to the right lead, or
the $\uparrow$ electron goes to the right lead first
(energy denominator $\eps$) and then the $\downarrow$
electron goes from the left lead to the dot. The amplitudes for
these two last contributions, which simply accounts for the two
different intermediate states, has to be added. Thus the total conductance
is given by:
\bq\label{G2}
G^{(0)}=G_Q\Gamma_L\Gamma_R
\left[
 \left(\fr{1}{\eps}-\fr{1}{\eps+U}\right)^2
 +\fr{1}{\eps^2}+\fr{1}{(\eps+U)^2}
\right]  .
\ee
The quantum dot in the regime described by  eq.~(\ref{G2})
contains one electron on the broad level $1$. Effective
antiferromagnetic interaction of this electron with the electrons
in the lead conduction bands, leads to a strong enhancement of
conductance at low temperature. This is the Kondo effect in
quantum dots~\cite{KondoT1,KondoT2,KondoT3}. With the use of
Schrieffer-Wolf transformation one easily maps the Anderson impurity model
Hamiltonian (eq.~(\ref{Ham}) with only one level) onto the Kondo
Hamiltonian (see e.g. \cite{Hewson}). Far from the resonances the
Kondo corrections to eq.~(\ref{G2}) are of the relative order
$\sim\ln(T^{-1})\Gamma/U$ where $T$ is a (small) temperature.
At $T\approx T_K=(U\Gamma/2)^{1/2}e^{\pi \eps_d(\eps_d+U)/2\Gamma U}$
the renormalised antiferromagnetic coupling diverges and the
conductance reaches the unitarity limit $G\approx G_Q$. Explicit
calculation of $G$ in this regime may be done only by means
of numerical renormalisation group. Still even in this
extreme case the eq.~(\ref{rising}) allows for qualitative description
of singular behavior of $G$.

Two kinds of sharp features are seen in fig.~2. First, there
are a cusp and a jump at the peaks $W^{I}$, where the curve
$G^{(1)}$ is replaced by $G^{(0)}$ and $W^{II}$, where
$G^{(2)}$ is replaced by $G^{(1)}$.
Accurate analytical description of $G(V)$ in this mixed valence
regime is possible only at $T\gg\Gamma$. (Although at such a
high temperature the singularity becomes smeared out, see the
next Section.)
The jump vanishes for
$\Delta\ll\Gamma$, but the pronounced cusp survives even for
$\Delta=0$.

Besides that, there are three jumps in the valleys.
The values of the conductance at $V=V^I\pm 0$ ($V^{III}\mp 0$) are
(assuming $\Delta \ll\Gamma$)
\bq\label{jump}
G\approx 48G_Q
\fr{\Gamma_L\Gamma_R}{U^2} \ \ {\rm and} \ \
G\approx 16G_Q
\fr{\Gamma_L\Gamma_R}{U^2} \ .
\ee
The contribution from the spin flip processes for $V<V^I$ is twice
that from the elastic processes. Therefore,
the conductance drops near $V=V^I$ by a factor of 3.
At $V=V^{II}$ the two electrons jump from the broad to
narrow level.
The discontinuity at $V^{II}$, which follows from the small
difference in the probability of the electron-like and hole-like
processes, vanishes for $\Delta\ll\Gamma$ (restoration of particle-hole
symmetry at $\Delta=0$).

In this paper we consider mainly the zero bias effects.
The finite bias conductance $G_b$
may also be found easily, assuming that the same model eq.~(\ref{Ham})
describes the quantum dot at finite bias (see however \cite{Levinson} for a discussion
of self-consistent screening, strictly valid for small bias only).
In the limit $\Gamma_{12},\Gamma_2=0$,
the differential conductance
$G_b$ acquires a cusp at the bias voltage $V_b$ coinciding with the energy
difference between the lowest and first excited state for the corresponding
valley. This singularity should be seen as two lines inside the Coulomb
blockade diamond relatively close to the zero bias diagonal ($V_b\ll U$)
crossing correspondingly at the gate voltage $V=V^I,V^{II},V^{III}$.
Namely this is $V_b=\pm|E^{(0)}-E^{(1)}|$ for the first valley,
$V_b=\pm|E^{(2)}-E^{(1)}|$ for the third valley and
$V_b=\pm|E^{(0)}-E^{(1)}|=\pm|E^{(2)}-E^{(1)}|$
for the second valley.
The explicit calculation of $G_b(V_b)$ may be done with the use of standard
master equation techniques.

\section{ Resolving sharp features.}

So far, we have considered only the case of an ``invisible'' second
level in the dot
$t_2^{L,R}\equiv 0$. However, even in this limit the
sharp features shown on fig.~2 should be smoothed due to the
finite temperature. We show this smoothening for the
cusp$+$jump at $V=W^I$ on the left panel of the fig~ 3.
The corresponding analytic expression
\bq
G=\fr{G^{(1)}\exp\{-E^{(1)}/T\}+G^{(0)}\exp\{-E^{(0)}/T\}}
{\exp\{-E^{(1)}/T\}+\exp\{-E^{(0)}/T\}}
 \ .
\ee
accounts simply for a different probability of thermal populations of
the states $E^{(1)}$ and $E^{(0)}$ of the dot. The same formula
describes the jump of the conductance in the middle of the
left ($V\approx V^I$) valley. Here the smoothed conductance has
a form of a Fermi function $(\exp\{ (E^{(0)}-E^{(1)})/T\}+1)^{-1}$.

\begin{figure}[t]
\epsfxsize=8.8cm
\epsffile{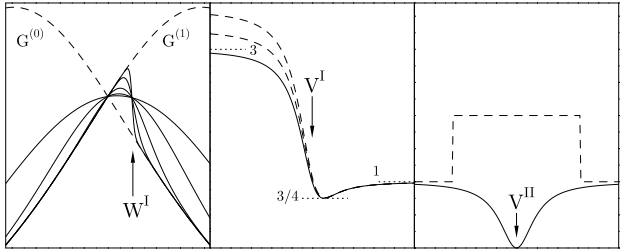}
\vglue 0.2cm
\medskip
\caption{The fine structure of the sharp features of
$G(V)$ seen in Fig.~2. {\it Left}: The smearing of the second
peak due to the finite temperature $T$ at few values of $T$.
Dashed lines are the conductances for the dot staying in the state
$(0)$ or $(1)$.
{\it Middle}: $G(V)$ at the middle of the first valley $V\approx V^I$,
without the Kondo effect (solid) and with Kondo effect included
for two different temperatures (eq. (18)). The numbers $1,3/4,3$ show
the relative
magnitude of the
current to the left and right of the transition and at the minimum
at $V=V^I+\gamma/\sqrt{3}$. {\it Right}: Conductance at the middle
of the second valley $V\approx V^{II}$ for the pure model eq.~(3)
(solid) and with the exchange interaction eq.~(21) added (dashed).
The scales are the same for middle and right figures.}
\end{figure}

The crossings of energy levels $E^{(i)}$ both at the peaks($W^i$)
and in the valleys($V^i$) become "avoided"due to the finite
coupling $t^{L,R}_2$ of the level 2 to the leads. This effect
determines the smearing of the conductance at
very low temperatures. Unfortunately we do not have a simple
way to take into account the coupling of the narrow level in
the mixed valence regime at $V\approx W^{I,II}$. On the other hand,
taking the second
coupling into account  at the valley, turns out to be relatively
easy~\cite{footnote}.
Consider for example the first crossing at $V\approx V^I$~(\ref{8}).
First, in a way analogous to eq.~(\ref{Gamma}) one may introduce
the width of the second level $\Gamma_2$ and the interlevel
width $\Gamma _{12}$
(defined as $\Gamma_{12} = 2\pi\sum_{i=L,R} t_1^i t_2^{i*}
{dn^i}/{d\eps}$).
In the first valley where is (almost) always one
electron in the dot either on the level 1, or on the level 2.
Hence we
may introduce the effective single-particle Hamiltonian,
accounting also for the coupling to the leads.
A simple  calculation in the second order of perturbation theory
gives for the elements of this Hamiltonian
\begin{eqnarray}\label{HamEff}
H_{11}&=&-\fr{\Gamma_1}{2\pi}\ln\left( \fr{\eps+U}{|\eps|}\right)
\, , \,
H_{22}=\Delta-\fr{\Gamma_2}{2\pi}\ln\left( \fr{\eps+U}{|\eps|}\right)
 , \nonumber\\
H_{12}&=&-\fr{\Gamma_{12}}{2\pi}\ln\left( \fr{\eps+U}{|\eps|}\right)
= H_{21}^* \  .
\end{eqnarray}
In particular here $H_{11}$ and $H_{22}$ are nothing more than the
renormalised single particle energies $\eps_1$ and $\eps_2$ given by
the first two formulae in eq.~(\ref{33}) (with $\Gamma_2$ included)
minus a proper constant (see also \cite{Haldane,Kaminski}).
Moreover, the small $\Gamma_2$ may
be omitted in the eq.~(\ref{HamEff}). Only  $\Gamma_{12}$
is important, since it is responsible for the mixing. We expect also
that $\Gamma_{12} \sim \sqrt{\Gamma_1\Gamma_2}
\gg\Gamma_2$.
Close to $V=V^I$, the logarithms in eq.~(\ref{HamEff}) may be
expanded in series. Now by simple diagonalization
of the $2\times 2$ matrix one finds the closest level splitting $\delta E$
and the width $\gamma$ (in gate voltage) of the avoided crossing.
\bq
\delta E=2\Delta \fr{\Gamma_{12}}{\Gamma} \ ; \
\gamma=\fr{\Gamma_{12}\Delta}{\Gamma^2} \pi U
\sim  \fr{\Delta}{\Gamma}\sqrt{\fr{\Gamma_2}{\Gamma_1}}U
\ .
\ee
The temperature is taken to be small compared to $\delta E$.

In the leading order coupling to the leads of the two levels
found after diagonalisation
of the $2\times 2$ Hamiltonian (\ref{HamEff}) is simply determined
by the amplitude of the broad level $1$ in the corresponding wave function
($\delta V=V-V^I$)
\bq\label{tpm}
t_{\pm}^{L,R}= t_{1}^{L,R} \sqrt{\fr{1}{2}\left(
1\pm \fr{\delta V}{\sqrt{\delta V^2 +\gamma^2}}
\right)} \ .
\ee
The calculation of the conductance with this $V$-dependent coupling
analogous to the derivation of the eq.~(\ref{G2}) gives
\begin{eqnarray}\label{smearstep}
G&=&2G_0\fr{\Gamma_L\Gamma_R}{(U/2)^2}\left(
1-\fr{\delta V}{\sqrt{\delta V^2+\gamma^2}}
+\fr{\delta V^2}{\delta V^2+\gamma^2} \right. \\
&+& \left. \left(
1-\fr{\delta V}{\sqrt{\delta V^2+\gamma^2}}
\right)^3 \fr{3\Gamma}{4\pi U}
\ln\left( \fr{U}{T}\right)
\right) \ . \nonumber
\end{eqnarray}
This result is illustrated in the central part of fig.~2.
The first line in the eq.~(\ref{smearstep}) is the result of
the calculation in the leading order of perturbation theory,
interpolating between the limiting values
(eq.~(\ref{jump})) at $\delta V \ll -\gamma$ and $\delta V\gg
\gamma$. In addition to the smearing of the step at $V\approx V^I$
the conductance eq.~(\ref{smearstep}) acquires a narrow minimum
at $\delta V=\gamma/\sqrt{3}$.

The last term in
eq.~(\ref{smearstep}) is the (first) Kondo correction. The Kondo
effect is present to the left of the drop of $G$ (where
the electron stays on the broad level), and vanishes (became
proportional to $\sim \Gamma_2^{\ 3}$) to the right of that drop.
Calculation of this low temperature Kondo correction is straightforward.
The factor $(1-\delta V/\sqrt{\delta V^2+\gamma^2})^3$ accounts for
the renormalised coupling of spin to the leads~(\ref{tpm}).
As usually this presence of the Kondo correction only at $V<V^I$ may be
easily checked by applying the small bias $V_b>T$.

The same effective Hamiltonian~(\ref{HamEff}) describes the
electronic configuration of the quantum dot in the second valley.
In this case one  simply  has always two electrons with opposite spins
occupying one of the two levels in the dot. The coupling of these two
levels to the leads changes rapidly at $V-V^{II}\sim \gamma$.
Since we always have a zero total spin, $S=0$, here,
there is no Kondo effect. The transmission
amplitude includes two competing contributions, electron-like
and hole-like, which lead to the vanishing of the conductance
at $V=V^{II}$ (see fig.~3, right frame)
\bq
G\sim \fr{(V-V^{II})^2}{(V-V^{II})^2+\gamma^2} \ .
\ee
This formula is valid for $\Delta\ll\Gamma$, but $G$ has a node
for $\Delta\sim\Gamma$ as well.

In the constant interaction model, eq.~(\ref{Ham}), the transition
at the second valley is a
triple crossing $E^{(0)}=E^{(1)}=E^{(2)}$.
This degeneracy is easily lifted if, e.g., in addition to the
direct Coulomb interaction~(\ref{Ham}) one introduces the exchange
interaction of the usual form
\bq
H_{exchange}=J\sum_{\sigma,\sigma'}
a^+_{1,\sigma} a_{2,\sigma}  a^+_{2,\sigma'} a_{1,\sigma'} \ .
\ee
As long as $J$ is sufficiently small ($J\ll\Gamma$) the level crossing at
$V=V^{II}$ will be splitted into two close crossings with a spin $1$
ground state of the quantum dot in between. The conductance in the three parts
of the valley is now proportional to
\begin{eqnarray}\label{JS1}
&&
\fr{2}{(\eps+U)^2} \ \, ; \\
&&
\fr{1}{(\eps+U)^2}+
\fr{1}{(\eps+2U)^2}+
\fr{1}{2}\left( \fr{1}{\eps+U}-\fr{1}{\eps+2U}\right)^2 \  ; \nonumber\\
&&
\fr{2}{(\eps+2U)^2} \ . \nonumber
\end{eqnarray}
The derivation of this result essentially repeats the proof of
eqs.~(\ref{G1},\ref{G2}). The second expression in (\ref{JS1})
corresponds to the
transmission through the dot having total spin $S=1$. Although only
one of the two electrons constituting this $S=1$ is visible
(i.e. well coupled to the leads) the
spin conservation
within transition leads to a factor $1/2$ in the spin-flip contribution.
Since in eq.~(\ref{JS1}) we describe the states with different spin,
the crossing of the eigenstates is not avoided and the transitions are
abrupt (at low $T$). Due to the eq.~(\ref{JS1}) the current is
enhanced around the transition in the second valley. The Kondo
effect (at $S=1$) leads to a further increase of the transmission.
The interesting physics of the Kondo effect at the triplet-singlet
transition
was investigated recently both experimentally~\cite{Sasaki} and
theoretically~\cite{Pustilnik1,Nazarov,Pustilnik2}.

\section{ Transmission through a multilevel dot ($\Gamma\gg
\Delta$).}

It is generally believed that the Coulomb blockade
in multilevel quantum dot may be seen only if the connection of the dot to the
leads is weak enough.
Namely one expects the parameters of the dot to be chosen to satisfy
the inequality
\bq\label{GDU}
\Gamma\ll\Delta< U \ ,
\ee
where $\Gamma$, $\Delta$ and $U$ are the typical width, the level
spacing and the charging energy. Just in order to fulfill this
condition and still to have a larger $\Gamma$, very small
quantum dots with large $\Delta$ were prepared for the Kondo
experiments~\cite{Kondo1,Kondo2,Kondo3}. However, in the case
when only one level
in the dot has an anomalously large width, the
inequality (\ref{GDU}) may be weakened to~\cite{SI}:
\bq\label{GU}
\Delta\ll \Gamma< U \ ,
\ee
where $\Gamma$ is the width of single anomalously broad
level. Still $\Gamma_i\ll\Delta$ for all other levels. Such rare
dominant levels are common for integrable dots and for dots
with the mixed (partly regular and partly chaotic) classical
dynamics. {\em Thus, instead of miniaturisation
of the quantum dot, one may try to look for the new physics by making the
dot more clean and symmetric}~\cite{Esteve}.

The model (\ref{GU}) with spinless electrons was used in
ref.~\cite{SI} in order to explain the transmission phase behavior
observed in the double slit experiment of the ref.~\cite{Heiblum2}.
Here we briefly discuss the modification of the same scenario
due to the spin.

In general for $\Gamma\gg\Delta$, charging of the quantum dot for a large
series of resonances resembles that for adding of the second and
third electron into the two level dot considered in the previous sections.
In the case of large $\Gamma$ the electron taken to the dot from the
leads within the charging resonance always occupies the broad level
(even if there are empty narrow levels with smaller single-particle
energies). This effect may be understood by introducing the $\sim\Gamma$
logarithmic correction to the single-particle energies in the dot
arising due to the coupling to the leads (see the eq.~(\ref{HamEff})).
Due to the spin degeneracy of the single-particle
levels a larger gain in energy is achieved if after the charging
resonance the broad level becomes occupied by two electrons.
Since only one external electron may be added to the dot at
the resonance, the second electron is taken to the broad level
from the narrow one (of course, if there is such electron with
close enough energy). This fast rearrangement of the electronic
configuration within the resonance leads to a new structure of the
conductance peaks (see two central peaks on the fig.~2). In the
middle of the valley both two electrons from the broad level
jumps to the narrow level (again if there is such an empty close
level). However, this double jump corresponds to the triple level
crossing of the states of the quantum dot~(\ref{33333}) and may be easily
splitted into two jumps by going beyond the constant interaction
model (fig.~3, right).

In \cite{SI} it was found that for $\Gamma \gg \Delta$ the
transmission phase $\phi$
through the quantum dot increases by $\pi$ through each charging peak
and sharply decreases by $\pi$ in $(2 \Gamma / \pi \Delta)
\ln( U / \Gamma )$ valleys.
In the limit of vanishing width
of all the narrow levels the function $\phi(V)$ may be constructed
simply from the function $\phi^{(0)}(V)$
describing the charging of single level. The procedure is similar
to the way we found the conductance
$G$ from the single level function $G^{(0)}(V)$ on fig.~2
with the use of eq.~(\ref{rising}).
In the spinless case the function $\phi^{(0)}(V)$ is described by
the Breit-Wigner formula. For the spin $S=1/2$ (the Anderson
impurity model) the $\phi^{(0)}(V)$ may be taken  e.g. from
ref.~\cite{Delft}. Compared to the case of the ref.~\cite{SI}
the main new effect
taking place due to spin is the sharp depopulation of the narrow level
at the charging resonance. The phase at the peak now first
increases smoothly from $\phi\approx 0$ to the value somewhat below
$\pi/2$, then jumps up by some fraction of $\pi$ and then continues
the smooth increase towards $\phi\approx \pi$. The finite temperature
tends to smear this three-stage increase of phase. The plateau with
$\phi\approx \pi/2$ at the Kondo valley predicted in ref.~\cite{Delft}
does not appear in the model (\ref{GU}). The crude behavior of the
transmission phase remains the same as in refs.~\cite{SI} (and
consistent with the experiment \cite{Heiblum2}). The phase increases
by $\pi$ at the resonance (although now the increase contains both
smooth and abrupt component) and drops down abruptly close to the
middle of the valley.

Thus taking into account spin does not lead to a serious revision
of the explanation~\cite{SI} of the transmission phase behaviour
of the experiment of ref.~\cite{Heiblum2}. The sequence of
resonances accompanied by the $-\pi$ jumps in the valley
is even doubled because of doubling of single particle density of
states due to spin.
Other attempts to explain the same experiment may be
found in ref.~\cite{other}. Although, none of the mechanisms presented
so far is capable to explain  by itself the observed phase behaviour.

\section{ Conclusions.}

In this paper we have considered the effects of the spin on transport through a
quantum dot having one level which is strongly  coupled to the leads. If this
coupling is strong enough, the coupled level
will play an essential role for the
energetics of the dot and it can change the distribution of
electrons over the discrete single-particle levels. We
introduced the model for multilevel quantum dot~(\ref{Ham},\ref{Gamma})
with one level having its Breit-Wigner width larger than the level
spacing $\Gamma>\Delta$. For spinless electrons
(and $\Gamma\gg\Delta$) the analogous
model was investigated in
our recent paper~\cite{SI}. Here we were mostly interested in the
effect of spin for electronic transport. Following the experimental
tendency towards  miniaturisation of the quantum dots we also mainly considered
the sequence of charging resonances corresponding to occupation of
only two levels by four electrons.

The main new effect taking place due to spin in our model is
the fast change of electronic configuration close
to the top of the charging resonance (in the mixed valence).
Within this transition one of the electrons in the dot
jumps from the narrow level to the well coupled one~\cite{Kulik}.
The population of narrow levels in our model
takes place in the Coulomb blockade valleys
and also leads to the peculiar behaviour of the conductance
(fig.~3).

Finally, due to a tendency to a double occupation(depopulation)
of well coupled levels we predict the suppression of the usual
(odd valley $S=1/2$) Kondo effect for $\Gamma >\Delta$. On the other hand,
the possibility to observe the new sharp features in the mixed-valence
regime, as well as $S=1$ Kondo effect and singlet-triplet transitions
in the valley, should partially compensate this drawback.

We acknowledge K.~Kikoin and D.~Esteve for valuable discussions.
This project was supported by the Israel Science Foundation, Jerusalem
by the German-Israeli Foundation (GIF)
and by a joint grant from the Israeli Ministry of Science
and the French Ministry of Research and Technology.
The work of P.G.S. was supported
by Dutch Science Foundation NWO/FOM.



\begin{thebibliography}{99}
\bibitem{Esteve}It has been remarked also to the authors by D.~Esteve
that the role of the narrow
levels storing the charge, might be played by different types of levels.
For example,
for strong enough metallic screening, the added electronic charge is
accumulated
near the gate, screening its potential. Thus, these effects may occur
quite generally
in metallic quantum dots.
\end{thebibliography}
\end{document}